\documentclass[fleqn,usenatbib]{mnras}

\usepackage[table]{xcolor}

\usepackage[T1]{fontenc}

\DeclareRobustCommand{\VAN}[3]{#2}
\let\VANthebibliography\thebibliography
\def\thebibliography{\DeclareRobustCommand{\VAN}[3]{##3}\VANthebibliography}

\usepackage{graphicx}	
\usepackage{amsmath}	
\usepackage{amssymb}	
\usepackage{newtxtext,newtxmath, color}

\definecolor{referee}{rgb}{0,0,0}
\def\changeref		{\color{referee}}

\title[Colour does not equal Morphology]{Quantifying the Poor Purity and Completeness of Morphological Samples Selected by Galaxy Colour}

\author[Galaxy Zoo team]{Rebecca J. Smethurst$^{1}$,\thanks{E-mail: rebecca.smethurst@physics.ox.ac.uk} Karen L. Masters$^{2}$,\thanks{E-mail: klmasters@haverford.edu}\thanks{Note that Smethurst \& Masters are both considered first authors of this paper} Brooke D. Simmons$^{3}$, Izzy L. Garland$^{3}$, Tobias G\'eron$^{1}$, \newauthor Boris H\"au{\ss}ler$^4$, Sandor Kruk$^{5,6}$, Chris J. Lintott$^{1}$,  David O'Ryan$^{3}$, Mike Walmsley$^{7}$
\\
$^{1}$Oxford Astrophysics, Department of Physics, University of Oxford, Denys Wilkinson Building, Keble Road, Oxford, OX1 3RH, UK\\ 
$^{2}$Departments of Physics and Astronomy, Haverford College, 370 Lancaster Avenue, Haverford, Pennsylvania 19041, USA\\ 
$^{3}$Physics Department, Lancaster University, Lancaster, LA1 4YB, UK\\
$^{4}$European Southern Observatory, Alonso de Cordova 3107, Vitacura, Santiago, Chile\\
$^{5}$Max-Planck-Institut f\"ur extraterrestrische Physik (MPE), Giessenbachstrasse 1, D-85748 Garching bei M\"unchen, Germany\\
$^{6}$European Space Agency, ESTEC, Keplerlaan 1, NL-2201 AZ, Noordwijk, The Netherlands\\
$^{7}$Department of Physics, The University of Manchester, Oxford Road, Manchester, M13 9PL, UK
}

\date{Accepted 2021 December 07. Received 2021 December 06; in original form 2021 August 06}

\pubyear{2021}

\begin{document}
\label{firstpage}
\pagerange{\pageref{firstpage}--\pageref{lastpage}}
\maketitle

\begin{abstract}
The galaxy population is strongly bimodal in both colour and morphology, and the two measures correlate strongly, with most blue galaxies being late-types (spirals) and most early-types, typically ellipticals, being red. This observation has led to the use of colour as a convenient selection criteria to make samples which are then labelled by morphology.  
Such use of colour as a proxy for morphology results in necessarily impure and incomplete samples. 
In this paper, we make use of the morphological labels produced by Galaxy Zoo to measure how incomplete and impure such samples are, considering optical ($ugriz$), NUV and NIR ($JHK$) bands. The best single colour optical selection is found using a threshold of $g-r = 0.742$, but this still results in a sample where only 56\% of red galaxies are smooth and 56\% of smooth galaxies are red. 
Use of the NUV gives some improvement over purely optical bands, particularly for late-types, but still results in low purity/completeness for early-types.
No significant improvement is found by adding NIR bands. With any two bands, including NUV, a sample of early-types with greater than two-thirds purity cannot be constructed. 
{\changeref Advances in quantitative galaxy morphologies have made colour-morphology proxy selections largely unnecessary going forward; where such assumptions are still required, we recommend studies carefully consider the implications of sample incompleteness/impurity.}
\end{abstract}







\section{Introduction}

{\changeref One of the central observations to be explained by any model of the galaxy population is that the dynamical histories and the star-formation histories of galaxies correlate well. Dynamical histories are imperfectly traced by galaxy morphology, and star-formation histories by integrated colours. The observation of a correlation between colour and morphology, alongside correlations of both properties with large scale galactic environment, are among the central pieces of evidence of models of hierarchical galaxy evolution, where massive galaxies assemble over time through multiple minor-, and occasional major-mergers \citep[e.g. as described in][]{SteinmetzNavarro2002}.}

{\changeref As astronomical surveys grew in size, obtaining visual morphology became more challenging.} The idea of making morphological selections using colour as a proxy for morphology {\changeref provided a convenient solution and became popular once} large scale, semi-automated imaging surveys (e.g. the Sloan Digital Sky Surveys, or SDSS, Main Galaxy Sample, MGS, \citealt{Strauss2002}) {\changeref quantified} the correlation between colour and morphology. Although it had been noted for decades before that that spiral galaxies tend to be bluer than elliptical galaxies (e.g. \citealt{Zwicky1955} comments that this had been ``known for a long time"), such large surveys confirmed the existence of the blue cloud dominated by disc galaxies and the red sequence dominated by elliptical galaxies \citep{Baldry2004, Baldry2006, Willmer2006, Ball2008, Brammer2009}. 

Many studies have {\changeref since gone on to} reveal the presence of significant fractions of spirals in the red sequence {\changeref ($\sim30$\%)} and/or smaller numbers of blue ellipticals {\changeref (up to $\sim10$\%; see for example} \citealt{vandenBergh1976,Bamford2009,Schawinski2009,Skibba2009,Wolf2009,Masters2010red,Bundy2010,Rowlands2012,Bonne2015,Fraser-McKelvie2016,Mahajan2020,Tuttle2020,Xu2021}). {\changeref Despite this, there remains a persistent idea in the literature that there are colour thresholds which can be used to make a clean morphological sample.} 

The publication of ``Color Separation of Galaxy Types in the Sloan Digital Sky Survey Imaging Data" \citep{Strateva2001} may perhaps be credited in large part to the {\changeref common use of} colour to separate samples by morphology. \citeauthor{Strateva2001} used SDSS photometry of almost 150,000 galaxies, finding a strong bimodality in $u-r$ colours, and used the morphologies of 267 galaxies in that sample to make the claim that red galaxies ``roughly correspond" to early-types (which they define as Sa, S0 and ellipticals), while blue galaxies correspond to the late-types (defined as Sb and Sc). 

{\changeref The impact of this conclusion can be traced via the citation trail through the literature. For example, \citet{Bell2004} state that the definition of an early-type should be redefined in terms of colour due to the results of \citet{Strateva2001}, \citet{Hogg2002} and \citet{Blanton2003}. Similarly, following on from the work of \citeauthor{Strateva2001}, \citet{ParkChoi2005} claim that a cut in a two-dimensional colour-colour parameter space is accurate enough to replace visual morphological classification. In addition,} \citet[][using \citealt{Strateva2001} as the primary reference]{Faber2007} state that ``early-type E/S0s populate a narrow red sequence that is separated from bluer, star-forming spirals by a shallow valley", going on to say ``not only does color sort galaxies cleanly into bins, it is also highly relevant to the emergence of the Hubble sequence."

This practice of equating colour and morphology is not limited to optical photometry. For example, \citet{Wright2010} and \citet{Jarrett2017} both label regions on a WISE FIR colour-colour plot by a mix of morphology and emission properties (e.g. labels include ``spirals", but also LINERS, or Low Ionization Nuclear Emission Regions). Similarly, spectra or Spectral Energy Distributions (SED) are also sometimes labeled by morphology, for example \citet{Benitez2004} label their SED collections by morphology. 

It {\changeref also} remains a common practice, particularly in studies which use galaxies as tracers of the large scale structure of the Universe, {\changeref to implicitly assume} that blue colours indicate {\changeref disc} structure (or late-type) and red colours indicate ellipticals {\changeref (or early-type; e.g. \citealt{Dawson2013} use red optical colours to select a sample of passively evolving early-type galaxies for use in the Baryon Oscillation Spectroscopic Survey of SDSS-III)}. {\changeref This is equivalent to assuming that} the star formation histories (as traced imperfectly by colours) and orbital motions (as traced imperfectly by morphology) of galaxies are uniquely connected ({\changeref for a selection of further examples of this see} e.g. \citealt{Bell2004,Weinmann2006,VandenBosch2008}, and \citealt{Chilingarian2012} who claim that by moving to NUV bands, clean morphological cuts can be made). For a more extended list of examples of publications which conflate colour and morphology in galaxy samples, see the Introduction of \citet{Masters2019spirals}\footnote{With thanks to the participants of the ‘Galaxy Zoo Literature Search Volunteers’ for finding examples.}. 

{\changeref The use of colour as a proxy for morphology certainly \emph{is} reasonable as a first-order, low-redshift approximation. It is also considerably easier to automate colour measurements than visual morphologies in large extragalactic surveys. These two factors have likely strongly motivated the field's widespread adoption of this practice. However, in this short article we argue that the impurity and incompleteness this introduces into modern analyses of galaxy evolution is often significant, can lead to incomplete and/or biased conclusions, and is no longer technologically necessary.} To demonstrate this, we make use of the {\it Galaxy Zoo 2} (GZ2) morphologies (based on citizen science inspection of SDSS images) along with SDSS, 2MASS and GALEX photometry {\changeref to quantify the purity and completeness of galaxy samples using colour as a proxy for morphology to split into} ``smooth" (aka early-types; this selection in GZ2 includes both elliptical and visually smooth S0 galaxies) or disc galaxies (GZ2 selected ``featured or disc" galaxies). We will quantify and explore the impurity and incompleteness involved in using a colour selection as a proxy for morphology selection and discuss the biases this assumption may introduce. We recommend that any study {\changeref which uses colour to morphologically categorise galaxies should be cognisant of the limitations discussed.}  

We describe the data sources and sample selection in Section \ref{sec:data}, and our method in Section~\ref{sec:method}. Our Results are shown in Section \ref{sec:results}, and we conclude in Section \ref{sec:conclusions}. In the rest of this work we adopt the Planck 2015 \citep{planck16} cosmological parameters with $(\Omega_m, \Omega_{\lambda}, h) = (0.31, 0.69, 0.68)$, where distances are needed to create physical units.

\section{Data and Sample Selection}\label{sec:data}

 \begin{figure}
	\includegraphics[width=\columnwidth]{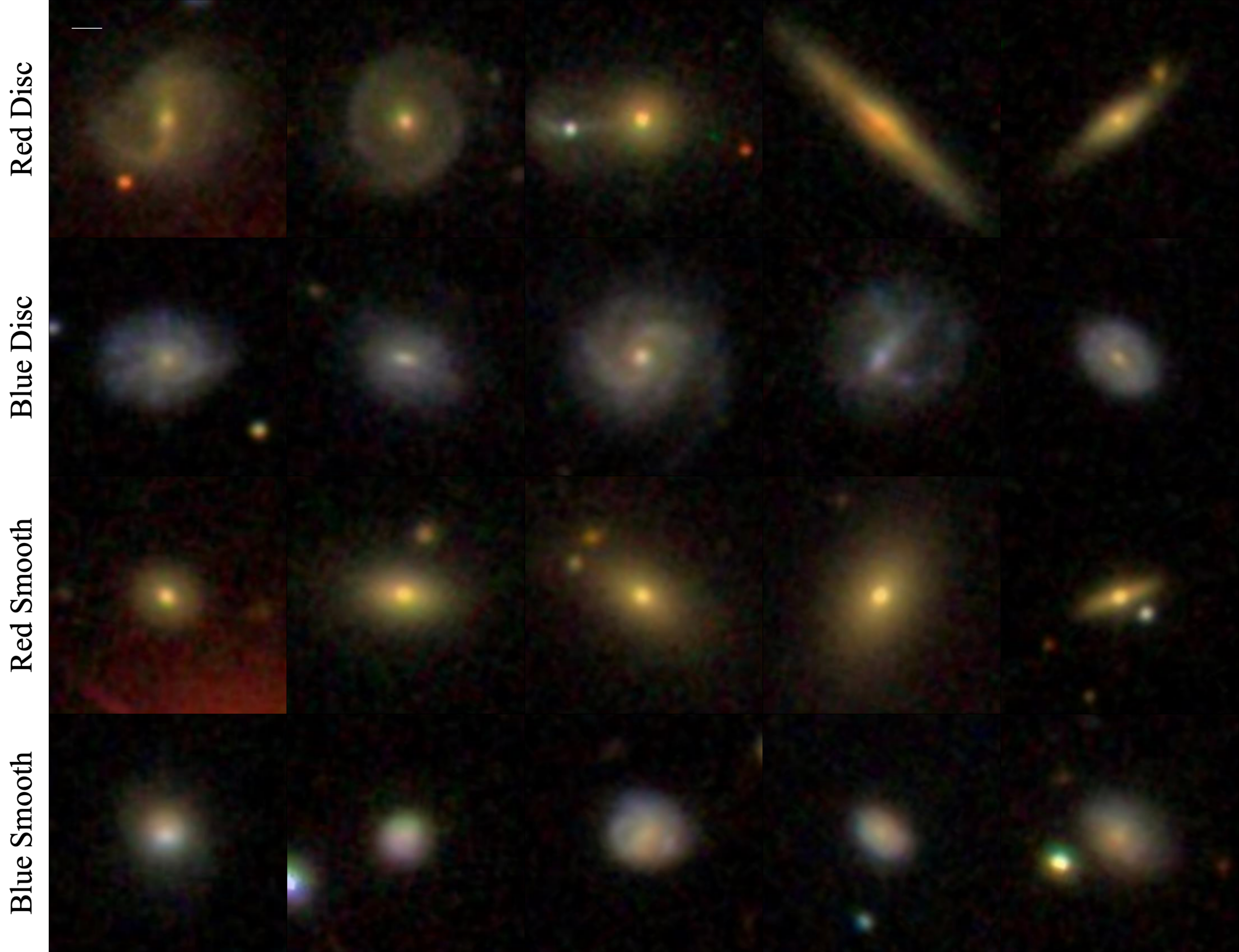}
    \caption{SDSS $gri$ postage stamp images showing 5 randomly chosen red discs (top row), blue discs (second row), red smooth (third row) and blue smooth galaxies (bottom row) in the redshift range $0.05 < z< 0.075$, morpholigcally classified using Galaxy Zoo vote fractions (see Section~\ref{sec:data}). The white bar in the top left panel shows the $5''$ pixel scale.}
\label{fig:mosaic}
\end{figure}

We used morphological classifications from {\it Galaxy Zoo 2} \citep{Willett2013} which were initially selected from the SDSS MGS  \citep{Strauss2002}, so have optical magnitudes available across $u$, $g$, $r$, $i$ and $z$ wavebands. This parent sample will be referred to as the \textsc{gz2sample} and has $239,695$ galaxies, corresponding to the brightest 25\% of the MGS (or $m_r<17$mag) from SDSS DR7 \citep{DR7} in the redshift range $0.01 < z < 0.24$ (median $z=0.075$). We cross-matched the \textsc{gz2sample} to the GALEX survey \citep{galex_paper} to obtain NUV magnitudes for $126,315$ galaxies matched with a search radius of $1''$ in right ascension and declination \citep[see][]{Smethurst2015}. This will be referred to as the \textsc{gz2galexsample} (median $z=0.068$). We also cross-matched the \textsc{gz2sample} to the 2MASS survey \citep{2mass_paper} to obtain $J$, $H$ and $K$ magnitudes for {\changeref $99,101$} galaxies. This will be referred to as the \textsc{gz22masssample} (median $z=0.077$). 
Finally, we also cross-matched the \textsc{gz2galexsample} to the \textsc{gz22masssample} to give {\changeref $99,065$} galaxies having optical, UV and NIR photometry, in the \textsc{gz2galex2masssample} (median $z=0.070$). 

{\changeref We used the SDSS Petrosian magnitudes, the GALEX {\tt auto} magnitudes and 2MASS XSC standard aperture (derived from the K$_s$-band 20 mag arcsec$^{-2}$ isophote, see Section 3.4 of \citealt{Jarrett2000}) to determine colours (for a discussion of aperture bias between different surveys see \citealt{Hill2011}).} All observed optical, ultraviolet and near-infrared magnitudes are corrected for galactic extinction \citep{Oh2011} by applying the \cite{Cardelli1989} law (giving a typical correction of $u-r\sim0.05$). We also adopt k-corrections to $z=0.0$ (following the method in \citealt{Bamford2009}). 

GZ2 morphological classifications are described in detail in \citet{Willett2013} but we briefly summarise here. GZ2 is a citizen science project which crowd sourced classifications from the public online. An average of around 40 volunteers classified each galaxy, using a tree of questions. For this work we focus only on the first question in the tree: ``Is the galaxy simply smooth and round with no sign of a disc?" to which the volunteers could answer "Smooth", "Features or disc" or "Star/artifact". Volunteer answers are aggregated into vote fractions after downweighting inconsistent answers. These vote fractions are then debiased to account for the impact of redshift on image quality. In this work we use the vote fractions debiased using the method described in \citet{Hart2016} which provides an improved debiasing technique over that initially presented in \citet{Willett2013}. Following this procedure we make use of the ``debiased vote fractions" for disc or featured galaxies ($p_d$) and smooth galaxies ($p_s$) to morphologically classify the galaxies in our samples.

We use a conservative cut to select a very pure sample of featured galaxies (most of which are spiral discs) as those with GZ2 debiased vote fractions of $p_d>0.8$ and smooth galaxies (ellipticals and featureless S0s) as those with $p_s>0.8$. Example SDSS images of those galaxies selected as disks and smooth are shown in Figure~\ref{fig:mosaic}. {\changeref Intermediate galaxies are not included in our analysis (those with $p_s\sim p_d\sim 0.5$), which are a mixture of genuinely intermediate galaxies (i.e. those of lenticular or S0 morphologies), and galaxies where morphological classification was inconclusive (i.e. due to poor image resolution or higher redshift)}. We note that detection fractions differ by morphology in our different subsets. For example, using the conservative cuts described above, in the \textsc{gz2sample}, 43\% of galaxies are discs, and 22\% smooth; this is very similar in the \textsc{gz22masssample}, but  changes to 41\% disc and just 8\% smooth galaxies in the \textsc{gz2galexsample} as smooth galaxies are more likely to be undetected in UV bands {\changeref  \citep{Smethurst2015, Schombert2016}}. {\changeref For a comparison of GZ2 morphologies to other morphological classification works, such as the expert visual classification of \cite{nair2010a} or the automatic classification of \cite{Huertas2011}, see Section 5 of \cite{Willett2013}.}


\section{Method}\label{sec:method}

In this work we want to investigate the purity and completeness of a sample when using colour as a proxy for morphology. In the literature, this is often done by applying a threshold cut in a chosen colour \citep{Bell2004, Faber2007, Weinmann2006, VandenBosch2008, Cooper2010, Zehavi2011, Ascasibar2011} to classify galaxies as either early-type/smooth (those redder than the threshold) or late-type/discs/featured (those bluer than the threshold). However, such samples will be contaminated by red late-types and blue early-types. Therefore to study the purity and completeness of a sample of early-type galaxies selected in this way we need to define:
\begin{itemize}
\item true positive (TP) = GZ smooth galaxies classified as early-type based on red colour (i.e. red early-types)
\item false positive (FP) = GZ featured/disc galaxies classified as early-types based on red colour (i.e. red spirals)
\item false negative (FN) = GZ smooth galaxies classified as late-type based on blue colour (i.e. blue early-types),
\end{itemize}

and similarly to select a sample of late-type (disc or featured) galaxies in this way:
\begin{itemize}
\item true positive (TP) = GZ featured/disc galaxies classified as late-types based on blue colour (i.e. blue spirals)
\item false positive (FP) = GZ smooth galaxies classified as late-type based on blue colour (i.e. blue early-types)
\item false negative (FN) = GZ featured/disc galaxies classified as early-type based on red colour (i.e. red spirals).
\end{itemize}

Then, at any given colour threshold, purity is the fraction of true identifications to all detections: 
\begin{equation}\label{eq:purity}
\rm{Purity}, P = \frac{TP}{TP+FP},
\end{equation}
and completeness is the fraction of true detections to all that should have been classified as true:
\begin{equation}\label{eq:completeness}
\rm{Completeness}, C= \frac{TP}{TP+FN}.
\end{equation}

We calculate both purity and completeness at 100 different colour values for all optical colour combinations in the \textsc{gz2sample}, all NUV-optical colour combinations in the \textsc{gz2galexsample}, all optical-infrared colour combinations in the \textsc{gz22masssample}, and all NUV-infrared colour combinations in the \textsc{gz22masgalexsample}. {\changeref The values were chosen as 100 equally spaced intervals between the 1st and 99th percentile of each colour distribution. We then determine the colour at which a compromise between purity and completeness is achieved for all magnitude band combinations, i.e. the colour at which the purity and completeness are equal.} 

\section{Results and Discussion}\label{sec:results}

\begin{figure}
	\includegraphics[width=0.985\columnwidth]{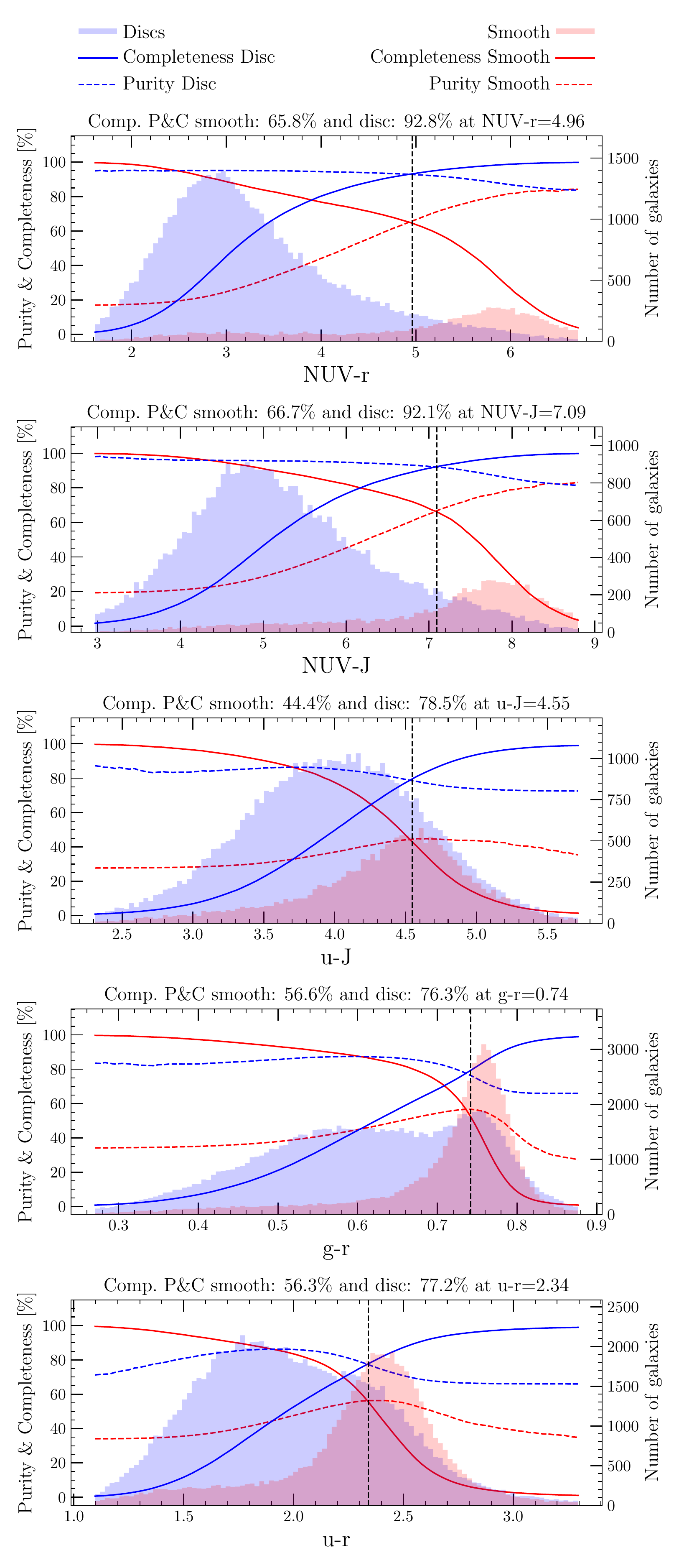}
	\vspace{-1em}
    \caption{The distribution of NUV$-r$ (top), NUV$-J$ (top middle), $u-J$ (middle), $g-r$ (bottom middle)  and $u-r$ (bottom) for galaxies classified as discs (blue histogram) and smooth (red histogram). The completeness (solid lines) and purity (dashed lines) are shown for selecting early-type (red lines) and late-types (blue lines) galaxy samples. These figures show how changing the colour threshold to morphologically classify galaxies based on their colour results in incomplete samples. {\changeref The black dashed vertical line shows the colour threshold at which a compromise between purity and completeness is achieved (i.e. they are equal)}. {\bf If there is no other option} but to use colour to make a morphological classification, then the purity and completeness are both maximised when a threshold of NUV$-r=4.96$, NUV$-J = 7.09$, $u-J=4.55$, $g-r=0.74$, or $u-r = 2.34$ is employed. }
\label{fig:coloursCP}
\end{figure}

\begin{table}
\centering
\caption{The compromised values between purity and completeness (i.e. when they are equal) achieved using colour to morphologically classify galaxies as either smooth or disc, with the corresponding colour threshold at which that compromise is achieved. Recommended colours with the highest values of compromised purity and completeness across UV, optical and infrared surveys are marked with an * and are shown in Figure~\ref{fig:coloursCP}.}
\begin{tabular}{cccc}
\hline
Colour & Comp. P/C Smooth [\%] & Comp. P/C Disc [\%] & Threshold \\
\hline

* NUV-r & 65.8 & 92.8 & 4.961 \\
NUV-i & 64.9 & 92.8 & 5.334 \\
NUV-z & 64.7 & 92.8 & 5.604 \\
NUV-g & 64.6 & 92.7 & 4.098 \\
NUV-u & 62.9 & 92.2 & 2.467 \\
* NUV-J & 66.7 & 92.1 & 7.091 \\
NUV-H & 66.9 & 91.8 & 7.716 \\
NUV-K & 65.6 & 91.6 & 8.286 \\
* u-J & 44.4 & 78.5 & 4.546 \\
u-H & 44.1 & 78.5 & 5.125 \\
u-i & 56.2 & 77.3 & 2.682 \\
* u-r & 56.3 & 77.2 & 2.340 \\
u-z & 56.2 & 76.9 & 2.947 \\
u-g & 55.2 & 76.8 & 1.601 \\
u-K & 40.0 & 76.8 & 5.745 \\
* g-r & 56.6 & 76.3 & 0.742 \\
g-z & 54.4 & 76.2 & 1.330 \\
g-i & 55.3 & 75.8 & 1.089 \\
g-H & 35.9 & 75.3 & 3.463 \\
r-i & 52.4 & 75.2 & 0.344 \\
g-J & 34.8 & 75.1 & 2.899 \\
g-K & 30.7 & 73.5 & 4.130 \\
r-H & 30.2 & 73.4 & 2.626 \\
r-z & 49.7 & 73.0 & 0.602 \\
r-J & 29.2 & 73.0 & 2.093 \\
i-H & 28.9 & 72.9 & 2.240 \\
i-J & 28.5 & 72.8 & 1.687 \\
r-K & 25.6 & 71.7 & 3.309 \\
i-K & 24.7 & 71.3 & 2.907 \\
z-H & 24.3 & 71.2 & 1.983 \\
z-J & 24.2 & 71.1 & 1.430 \\
z-K & 21.9 & 70.3 & 2.661 \\
i-z & 42.2 & 69.9 & 0.259 \\
\hline
\label{table:colourresults}
\end{tabular}
\end{table}

We investigate the purity and completeness of using a single colour threshold to select a morphological sample for various colour combinations across the UV, optical and infrared wavelength ranges of GALEX, SDSS and 2MASS. The colour threshold at which a compromise between purity and completeness is achieved (i.e. when they are equal) are achieved are listed in Table~\ref{table:colourresults}. 

We find that NUV$-r$ is the best colour to use to achieve an ideal compromise purity and completeness of $65.8\%$ for a sample of smooth galaxies and $92.8\%$ for disc galaxies, using a colour threshold of NUV$-r=4.961$. If optical observations are not available, then NUV$-J$ is the next best option with {\changeref $66.7\%$ purity/completeness for smooth galaxies and $92.1\%$ purity/completeness for disc galaxies at a colour threshold of NUV$-J = 7.091$}. If UV observations are not available then $u-J$ is the next best option with {\changeref $44.4\%$ purity/completeness for smooth galaxies and $78.5\%$ purity/completeness for disc galaxies at a colour threshold of $u-J=4.290$}. If neither UV or infrared observations are available the next best option is  $g-r$ with $56.6\%$ purity/completeness for smooth galaxies and $76.0\%$ purity/completeness for disc galaxies at a colour threshold of $g-r=0.742$ (similarly, $u-r$ also achieves a purity/completeness of $56.3\%$/$77.2\%$ at a colour threshold of $u-r=2.340$). {\changeref In Figure~\ref{fig:coloursCP} we show how the purity and completeness change with different colour thresholds for each of these five colours, demonstrating how our quoted thresholds in Table~\ref{table:colourresults} result in a compromise between maximising purity and completeness.}

The addition of NIR bands to optical bands does not significantly improve the compromise between purity and completeness achievable when only optical observations are available. However, it is clear from Table~\ref{table:colourresults} that having rest frame NUV magnitudes does significantly improve the compromise between purity and completeness when using colour to morphologically classify a sample, especially for constructing a disc galaxy sample. In this we agree with the findings of \citet{Chilingarian2012} who suggest NUV-$r$ is a significantly better choice than optical selection when dividing the galaxy population. They suggest a threshold of NUV$-r>4$ to select early-types, as opposed to the NUV$-r=4.961$ threshold we found in this study (see Table~\ref{table:colourresults}). Using the \cite{Chilingarian2012} cut of NUV$-r>4$, results in a sample of early-type galaxies with a purity of $44.1\%$ and a completeness of $76.8\%$, and a late-type sample with a purity $94.5\%$ and a completeness of $80.5\%$. Therefore $\sim20\%$ of disc galaxies will be missed using the \cite{Chilingarian2012} threshold, and these missing disc galaxies consist of ``red spiral" galaxies of particular interest for understanding quenching of star formation {\changeref \citep{Masters2010red, Cortese2012, Tojeiro2013, Fraser-McKelvie2016, Fraser-McKelvie2018, Mahajan2020}}.

{\changeref However, as is clear from comparing the red-filled distributions across the panels of Figure~\ref{fig:coloursCP}, a significant number of smooth galaxies are not detected in NUV bands. This reduces the ultimate sample size and the effectiveness of NUV colours at selecting a sample of early-type galaxies. Unfortunately}, one cannot assume that all GALEX non-detections are smooth galaxies; of the non-detections in the \textsc{gz2sample} that are in the GALEX footprint \citep[using the GALEX–SDSS–WISE Legacy catalog of][]{salim16}, $35\%$ are discs and $29\%$ are early-type galaxies. {\changeref Therefore, GALEX non-detections do not improve the morphological selection of smooth galaxies. In addition,} as the bottom panels of Figure~\ref{fig:coloursCP} show, optical magnitudes alone do not allow for an accurate morphological classification of a sample of smooth galaxies by colour. This is particularly apparent for $u-r$ and will affect all such works that use $u-r$ colours to split the galaxy population morphologically. For example, \citet{Strateva2001} use a colour threshold of $u-r=2.22$, which is similar to the threshold found in this study of $u-r=2.340$ (see Table~\ref{table:colourresults}) which results in $\sim44\%$ of galaxies classified as early-type based on a red colour actually having featured disc morphologies.


 In this work we only assess the purity and completeness of single colour cuts. {\changeref However, it is common to use a magnitude dependent cut to divide the galaxy population. \citet{Baldry2004} fit an optimal division of $u-r = 2.06 - \tanh{(M_r+20.07)/1.09}$ which varies from, $u-r=1.8$ at the faint end, to $2.3$ at the bright end. They caution that while this is an optimal divider, significant overlap exists in the two populations they model as Gaussian in colour.} Using this division with the \textsc{gz2sample} results in a sample of early-type galaxies with a purity of $56.8\%$ and a completeness of $70.5\%$, and a late-type sample with a purity $82.6\%$ and a completeness of $72.3\%$. Comparing these numbers found using the \citeauthor{Baldry2004} division to those for a single $u-r$ colour selection stated in Table~\ref{table:colourresults}  reveals that the only significant improvement is to the completeness of the sample of early-type galaxies selected, since the two-dimensional \citeauthor{Baldry2004} threshold moves bluer than our selection for fainter galaxies.  Similarly, \citet{Masters2010red} in identifying optically red spirals, use a magnitude dependent cut of $(g-r)=0.63 - 0.02(M_r +20)$, which was 1$\sigma$ bluer than the main ridge of the red sequence, which revealed up to 30\% (at the most massive end) of even the most face-on spirals are clearly in the red sequence.

{\changeref Along with colour-magnitude, the use of two colours to divide the galaxy population is also common. Examples include UV-optical combinations (e.g. NUV-$r$ against $g-r$; \citealt{Chilingarian2012}) optical-NIR combinations (e.g. the $UVJ$ diagram of $V-J$ against $U-V$; \citealt{Patel2012, Muzzin2013, Fang2018}). Typically, the main purpose of these 2-colour diagrams is to separate passively evolving galaxies from star-forming galaxies with dust-reddened optical colours. Given that sub-populations of interest such as red spirals are intrinsically red/passive rather than simply dust-reddened \citep{Masters2010red}, these distinctions have limited effect on improving the use of colour as a proxy for morphological selection. For example, using a two dimensional colour cut in NUV$-u$ against $u-r$ roughly reduces to a single colour cut in NUV$-r$, the colour we have shown is the best performing single colour cut in this study (see Table~\ref{table:colourresults}). Similarly, a two dimensional colour cut on the $UVJ$ diagram is still highly incomplete/impure if the cut between ``active'' and ``passive'' from that diagram is assumed to map to late-type and early-type morphologies, respectively. 

In fact, given that $UVJ$ selection is common in higher-redshift studies, the use of this 2-colour selection demonstrates how issues with conflating star-formation status and morphology can be exacerbated at higher redshift. If we combine \emph{Galaxy Zoo: Hubble} classifications \citep{Willett2017} with rest-frame $UVJ$ colours and stellar masses determined by COSMOS-Ultravista \citep{Muzzin2013}, we may examine purity and completeness using ``clean'', debiased samples of smooth and featured galaxies analogous to the selections described in Section \ref{sec:data} and within a volume limit of $z \leq 1$ (chosen to ensure morphologies are determined on the basis of rest-frame optical imaging) and stellar mass $M_\ast \geq 10^9\, $M$_\odot$. We find that assuming ``passive'' equates to early-type results in an early-type purity of 55\% and completeness of 48\%. The late-type assumption results in considerably higher purity (88\%) and completeness (91\%), but this is dominated by the fact that at higher redshifts featured galaxies are common: if we were to assume that \emph{all} galaxies were late-type regardless of colour in this ``clean'' subsample, the selection would still be 81\% pure. The exact numbers depend somewhat on the lower mass limit, but the qualitative result does not change for any reasonable choice of mass cut above the flux limit. Using rest-frame optical morphologies out to $z \sim 2$, an examination of $UVJ$ versus morphology by \citet{Simmons2017} reveals a similar mixing of morphologies across the passive/active $UVJ$ boundary, with an even larger fraction of smooth galaxies showing active star formation. We note that the practice of equating colour with morphology is much less common in high-redshift studies \citep[e.g., see the introduction of][]{Schreiber2018}, in part because the disconnect between these properties is more obvious at earlier epochs. The evolution of galaxy structures combined with the relatively poor mapping of colour to morphology strongly suggests there is little to be gained in the purity or completeness of a morphological sample by using a 2-dimensional colour-colour or colour-magnitude selection, at any redshift.}


 
 Similarly, other morphological proxies such as FracDev, $f_{\rm{Dev}}$, and S\'ersic index, $n$, also result in impure and/or incomplete morphological samples. For example \cite{Masters2010dust} demonstrate that $45\%$ of ``early types'' found by $f_{\rm{DeV}} > 0.5$ \cite[the FracDev cut used by][]{Strateva2001} are identified by Galaxy Zoo as featured discs, while $5\%$ of the ``spirals'' found by $f_{\rm{DeV}} < 0.5$ are identified by Galaxy Zoo as smooth galaxies. Use of a single-S\'ersic parameter as a proxy for bulge strength can also be challenging due to the fact that galaxies with ``intermediate’’ S\'ersic values (e.g., $1.5 < n < 3$) can be either bulge-dominated \emph{or} disk-dominated \citep[][H\"au{\ss}ler et al, in preparation]{Simmons2008}. Additionally, \citet{Lange2015} find that a S\'ersic-index selection is the least reliable parameter for discriminating between morphological early- and late-type galaxies, compared to $u-r$ and $g-i$ colour cuts. Some mitigations are possible: \cite{Vika2015} found using the \emph{ratio} of S\'ersic indices measured in two different filters could recover half of the disc galaxies erroneously classified as early-type galaxies by a joint $u-r$ and single band S\'ersic index selection. Morphological studies using two S\'ersic parameters to simultaneously characterise bulge and disk can be more reliable than single-S\'ersic fits provided there is adequate signal-to-noise in the images \citep{Simard2011}, and in some cases a joint selection using both Galaxy Zoo and S\'ersic-based bulge-to-total morphologies allow for selection of more pure and complete samples \citep{Simmons2017}. A joint selection capitalises on the fact that Galaxy Zoo’s top-level ``smooth or featured’’ question does \emph{not} fundamentally capture a ``S\'ersic by eye’’. One additional complication is that in strongly barred disc galaxies, the presence of the bar can bias the fit making the bulge component appear larger and more concentrated (higher S\'ersic index) than it is \citep{Kruk2018}.

As an alternative to the colour selection, an increasing body of work relies on machine learning to provide classifications of galaxies {\changeref \citep[e.g.][]{Huertas2015, Beck2018, dominguezsanchez2018, Walmsley2020, Vavilova2021}}. Though early examples date from nearly thirty years ago \citep{storrie92} the development of neural networks was a significant breakthrough. The first application to galaxy classification by \cite{Ball2004} found that the highest values of correlation between T-type and input parameters, including properties such as radius, surface brightness and concentration, were with colour, an early indication of the propensity of the correlation between colour and morphology to influence machine learning classifications. This property is particularly seen in more recent deep learning implementation by, amongst others, \cite{Dieleman2015, dominguezsanchez2018, dominguezsanchez2019, Vavilova2021}. {\changeref \citet{Vavilova2021} claim that their supervised machine learning algorithm trained over SDSS photometric parameters is less biased than when trained using Galaxy Zoo visual morphologies. However, their measurement of bias is based on the correlation of the morphologies with colour and concentration, exactly the assumption we are concerned to avoid making here.}

The correlation between colour and morphology make it a confounding variable in machine learning solutions; without training on monochrome images or the development of a specialised figure of merit, any supervised network is likely to quickly learn the apparent rule that colour implies morphology. For example, \cite{hocking15} used an unsupervised learning technique to split late- and early-types in cluster survey images; again with colour being the dominant property that the machine used to classify. \cite{hocking15} discuss how galaxies identified as early-types by their algorithm include those spirals with bulges with redder colours, and those galaxies identified as late-types included lensed features with bluer colours. Such careful reflection on results demonstrates the issues that arise when an algorithm uses colour as a proxy for morphology. 
 

{\changeref On the other hand, when morphology and colour are used independently as (imperfect) measures of dynamics and star formation, new results often emerge. For example, \citet{Nair2010b} find that colour and morphology have different effects on both the slope and dispersion of local luminosity-size relations for galaxies. Investigating quenching, \citet{Schawinski2014} separate the local (SDSS) colour-magnitude diagram by visual morphology (using the same ``clean'' samples that we have used here) and find that the green valley can be understood as the overlap of the tails of the two morphological populations. Incorporating a simple model of star formation history using NUV-optical colours, \citeauthor{Schawinski2014} also find that discs and spheroids evolve very differently across the green valley. With Bayesian modelling of star formation histories from galaxy colours and incorporating the full galaxy population (including intermediate morphologies) using \emph{Galaxy Zoo} morphologies to provide probabilistic weighting, \citet{Smethurst2015} uncover further nuance in the quenching histories of smooth and featured galaxies (e.g. revealing an intermediate quenching mode which can occur both with and without morphological transformation), which would have remained hidden had colour been used as a proxy for morphology. In an examination of galactic conformity that (for the first time in this field; c.f. \citealt{Weinmann2006}; \citealt{Prescott2011}) considers morphology and star formation fully independently, \citet{Otter2020} find that star-formation and morphological conformity are \emph{different}. Specifically, they find that morphological conformity is weaker than star-formation conformity, which is consistent with a physical model wherein star formation properties change more rapidly (or more readily) than galaxy dynamics in the group environment. Another example is the discovery that red discs host significantly more bars, which \citet{Masters2011bars} argue had been previously missed in part due to pre-selection of ``disc" samples to include only blue discs. Results such as these offer a glimpse into the potential of independently examining morphology and star formation rate, in addition to the results on passive red spirals and blue ellipticals already discussed above.}

 The wide availability of Galaxy Zoo morphologies for a variety of imaging surveys\footnote{e.g. for GZ analysis of SDSS, DECaLS and various public HST surveys, see \url{https://data.galaxyzoo.org}} means that any selection a researcher decides must be made using colour can be checked for purity and completeness with quantitative visual morphologies. 
 {\changeref For most investigations into galaxy evolution where both colour and morphology are involved, we would strongly suggest using morphology as an independent quantity wherever possible. If quantitative visual morphologies are not available for a particular sample, a combination of parametric and/or non-parametric morphologies may be used (potentially in combination with colour) to increase a sample's purity and completeness. }

 %


\section{Conclusions}\label{sec:conclusions}

We have investigated the purity and completeness {\changeref of galaxy samples constructed using a single colour cut as a proxy for morphology}. We determined the values of purity and completeness that can be achieved using a given colour threshold for colours across optical (SDSS; $ugriz$), ultraviolet (GALEX; NUV) and infrared (2MASS; $JHK$) surveys. {\changeref We focus only on single colour cuts in this study (however see Section~\ref{sec:results} for a discussion on the addition of magnitude, a second colour or other morphological proxies) and determine the colour threshold at which 
a reasonable 
compromise between purity and completeness can be achieved. 
We choose to examine the value where purity and completeness are equal, as this single value being higher typically indicates a higher fidelity in the assumption that colour and morphology are good proxies of one another.
}

{\changeref We find that $NUV-r$ achieves the best compromise between morphological purity and completeness within a sample. Using a colour threshold of NUV$-r=4.961$ results in a sample of early-type galaxies with a purity and completeness of $65.8\%$, and for late-type galaxies of 92.8\%.}

{\changeref Without the addition of NUV magnitudes, optical colours used as a proxy for morphology result in less pure and less complete samples}. Using a threshold of $g-r=0.742$ results in an early-type galaxy sample with a purity/completeness of only $56.6\%$, and a late-type galaxy sample with purity/completeness of $76.3\%$. We also find that no improvement is found by adding NIR magnitudes to the optical bands. {\changeref We note that with any two magnitude bands, including NUV, a sample of early-types with greater than two-thirds purity cannot be constructed.}
 
 {\changeref We therefore conclude that when colour is used as a proxy for morphology impure and incomplete samples are the result.} If no other option beyond a colour cut is available, either (1) samples should not be interpreted as morphologically homogeneous, or (2) the morphological make-up of colour selected samples should be measured, e.g. using publicly available Galaxy Zoo morphologies$^2$. 
  
 {\changeref The 
 relative
 simplicity of quantifying colour in large surveys, 
 versus 
 the complexity of morphology, presumably contributed to the large scale uptake of conflating colour with morphology. While this 
 has
 allowed for many significant 
 advances in our understanding
 of the general galaxy population, 
 further progress requires that we 
 more explicitly separate stellar dynamics from star formation by 
 incorporating
 the visual morphology of galaxies 
 into our analyses.
 }  The problem of generating high quality, quantitative (i.e. having an estimate of error) visual morphologies for large samples of galaxies was solved over a decade ago with the {\it Galaxy Zoo} methodology \citep{Lintott2008}, which is now being extended to work for larger and larger samples by an optimal partnership between the crowd and machines using adaptive learning \citep{Walmsley2021}. {\changeref Similarly, supervised and unsupervised machine learning have advanced in recent decades to produce purer and more complete morphological samples than using colour alone. In addition, non-parametric morphological classifications such as concentration and asymmetry \citep[e.g.,][]{Abraham1994,Abraham1996,Tohill2021} and the Gini coefficient and $M_{20}$ \citep{Lotz2004}, have provided an automated way of morphologically constraining large samples of galaxies.} Therefore, the main reasons that the community moved to widespread use of colour as a proxy for morphology within the galaxy population are 
 no longer valid. 
 {\changeref With upcoming large-scale extragalactic surveys poised to deliver a variety of robust morphological measures alongside precise multi-band photometry, we anticipate new discoveries about galaxy evolution from the careful consideration of these independent quantities.}

\section*{Acknowledgements}

IG and DO'R acknowledge support from STFC studentships at Lancaster University [grant number ST/T506205/1].
BDS acknowledges support from a UKRI Future Leaders Fellowship, administered by the Medical Research Council [grant number MR/T044136/1].

This publication uses data generated via the \url{Zooniverse.org} platform, development of which is funded by generous support, including a Global Impact Award from Google, and by a grant from the Alfred P. Sloan Foundation. This publication has been made possible by the participation of hundreds of thousands of volunteers in the {\it Galaxy Zoo} project on \url{Zooniverse.org.}

This publication makes use of data products from the Two Micron All Sky Survey, which is a joint project of the University of Massachusetts and the Infrared Processing and Analysis Center/California Institute of Technology, funded by the National Aeronautics and Space Administration and the National Science Foundation. This publication makes of use data from GALEX. GALEX is a NASA mission managed by the Jet Propulsion Laboratory. This publication uses data from the Sloan Digital Sky Survey (SDSS-I/II). Funding for the SDSS and SDSS-II has been provided by the Alfred P. Sloan Foundation, the Participating Institutions, the National Science Foundation, the U.S. Department of Energy, the National Aeronautics and Space Administration, the Japanese Monbukagakusho, the Max Planck Society, and the Higher Education Funding Council for England. The SDSS Web Site is http://www.sdss.org/.


\section*{Data Availability}

All data used in this paper is publicly available at the locations we cite.  
 



\bibliographystyle{mnras}
\bibliography{references.bib} 

\begin{thebibliography}{}
\makeatletter
\relax
\def\mn@urlcharsother{\let\do\@makeother \do\$\do\&\do\#\do\^\do\_\do\%\do\~}
\def\mn@doi{\begingroup\mn@urlcharsother \@ifnextchar [ {\mn@doi@}
  {\mn@doi@[]}}
\def\mn@doi@[#1]#2{\def\@tempa{#1}\ifx\@tempa\@empty \href
  {http://dx.doi.org/#2} {doi:#2}\else \href {http://dx.doi.org/#2} {#1}\fi
  \endgroup}
\def\mn@eprint#1#2{\mn@eprint@#1:#2::\@nil}
\def\mn@eprint@arXiv#1{\href {http://arxiv.org/abs/#1} {{\tt arXiv:#1}}}
\def\mn@eprint@dblp#1{\href {http://dblp.uni-trier.de/rec/bibtex/#1.xml}
  {dblp:#1}}
\def\mn@eprint@#1:#2:#3:#4\@nil{\def\@tempa {#1}\def\@tempb {#2}\def\@tempc
  {#3}\ifx \@tempc \@empty \let \@tempc \@tempb \let \@tempb \@tempa \fi \ifx
  \@tempb \@empty \def\@tempb {arXiv}\fi \@ifundefined
  {mn@eprint@\@tempb}{\@tempb:\@tempc}{\expandafter \expandafter \csname
  mn@eprint@\@tempb\endcsname \expandafter{\@tempc}}}

\bibitem[\protect\citeauthoryear{{Abazajian} et~al.,}{{Abazajian}
  et~al.}{2009}]{DR7}
{Abazajian} K.~N.,  et~al., 2009, \mn@doi [\apjs]
  {10.1088/0067-0049/182/2/543}, \href
  {http://adsabs.harvard.edu/abs/2009ApJS..182..543A} {182, 543}

\bibitem[\protect\citeauthoryear{{Abraham}, {Valdes}, {Yee}  \& {van den
  Bergh}}{{Abraham} et~al.}{1994}]{Abraham1994}
{Abraham} R.~G.,  {Valdes} F.,  {Yee} H.~K.~C.,   {van den Bergh} S.,  1994,
  \mn@doi [\apj] {10.1086/174550}, \href
  {https://ui.adsabs.harvard.edu/abs/1994ApJ...432...75A} {432, 75}

\bibitem[\protect\citeauthoryear{{Abraham}, {van den Bergh}, {Glazebrook},
  {Ellis}, {Santiago}, {Surma}  \& {Griffiths}}{{Abraham}
  et~al.}{1996}]{Abraham1996}
{Abraham} R.~G.,  {van den Bergh} S.,  {Glazebrook} K.,  {Ellis} R.~S.,
  {Santiago} B.~X.,  {Surma} P.,   {Griffiths} R.~E.,  1996, \mn@doi [\apjs]
  {10.1086/192352}, \href
  {https://ui.adsabs.harvard.edu/abs/1996ApJS..107....1A} {107, 1}

\bibitem[\protect\citeauthoryear{{Ascasibar} \& {S{\'a}nchez
  Almeida}}{{Ascasibar} \& {S{\'a}nchez Almeida}}{2011}]{Ascasibar2011}
{Ascasibar} Y.,  {S{\'a}nchez Almeida} J.,  2011, \mn@doi [\mnras]
  {10.1111/j.1365-2966.2011.18869.x}, \href
  {https://ui.adsabs.harvard.edu/abs/2011MNRAS.415.2417A} {415, 2417}

\bibitem[\protect\citeauthoryear{{Baldry}, {Glazebrook}, {Brinkmann},
  {Ivezi{\'c}}, {Lupton}, {Nichol}  \& {Szalay}}{{Baldry}
  et~al.}{2004}]{Baldry2004}
{Baldry} I.~K.,  {Glazebrook} K.,  {Brinkmann} J.,  {Ivezi{\'c}} {\v{Z}}.,
  {Lupton} R.~H.,  {Nichol} R.~C.,   {Szalay} A.~S.,  2004, \mn@doi [\apj]
  {10.1086/380092}, \href
  {https://ui.adsabs.harvard.edu/abs/2004ApJ...600..681B} {600, 681}

\bibitem[\protect\citeauthoryear{{Baldry}, {Balogh}, {Bower}, {Glazebrook},
  {Nichol}, {Bamford}  \& {Budavari}}{{Baldry} et~al.}{2006}]{Baldry2006}
{Baldry} I.~K.,  {Balogh} M.~L.,  {Bower} R.~G.,  {Glazebrook} K.,  {Nichol}
  R.~C.,  {Bamford} S.~P.,   {Budavari} T.,  2006, \mn@doi [\mnras]
  {10.1111/j.1365-2966.2006.11081.x}, \href
  {https://ui.adsabs.harvard.edu/abs/2006MNRAS.373..469B} {373, 469}

\bibitem[\protect\citeauthoryear{{Ball}, {Loveday}, {Fukugita}, {Nakamura},
  {Okamura}, {Brinkmann}  \& {Brunner}}{{Ball} et~al.}{2004}]{Ball2004}
{Ball} N.~M.,  {Loveday} J.,  {Fukugita} M.,  {Nakamura} O.,  {Okamura} S.,
  {Brinkmann} J.,   {Brunner} R.~J.,  2004, \mn@doi [\mnras]
  {10.1111/j.1365-2966.2004.07429.x}, \href
  {https://ui.adsabs.harvard.edu/abs/2004MNRAS.348.1038B} {348, 1038}

\bibitem[\protect\citeauthoryear{{Ball}, {Loveday}  \& {Brunner}}{{Ball}
  et~al.}{2008}]{Ball2008}
{Ball} N.~M.,  {Loveday} J.,   {Brunner} R.~J.,  2008, \mn@doi [\mnras]
  {10.1111/j.1365-2966.2007.12627.x}, \href
  {https://ui.adsabs.harvard.edu/abs/2008MNRAS.383..907B} {383, 907}

\bibitem[\protect\citeauthoryear{{Bamford} et~al.,}{{Bamford}
  et~al.}{2009}]{Bamford2009}
{Bamford} S.~P.,  et~al., 2009, \mn@doi [\mnras]
  {10.1111/j.1365-2966.2008.14252.x}, \href
  {https://ui.adsabs.harvard.edu/abs/2009MNRAS.393.1324B} {393, 1324}

\bibitem[\protect\citeauthoryear{{Beck} et~al.,}{{Beck}
  et~al.}{2018}]{Beck2018}
{Beck} M.~R.,  et~al., 2018, \mn@doi [\mnras] {10.1093/mnras/sty503}, \href
  {https://ui.adsabs.harvard.edu/abs/2018MNRAS.476.5516B} {476, 5516}

\bibitem[\protect\citeauthoryear{{Bell} et~al.,}{{Bell}
  et~al.}{2004}]{Bell2004}
{Bell} E.~F.,  et~al., 2004, \mn@doi [\apj] {10.1086/420778}, \href
  {https://ui.adsabs.harvard.edu/abs/2004ApJ...608..752B} {608, 752}

\bibitem[\protect\citeauthoryear{{Ben{\'\i}tez} et~al.,}{{Ben{\'\i}tez}
  et~al.}{2004}]{Benitez2004}
{Ben{\'\i}tez} N.,  et~al., 2004, \mn@doi [\apjs] {10.1086/380120}, \href
  {https://ui.adsabs.harvard.edu/abs/2004ApJS..150....1B} {150, 1}

\bibitem[\protect\citeauthoryear{{Blanton} et~al.,}{{Blanton}
  et~al.}{2003}]{Blanton2003}
{Blanton} M.~R.,  et~al., 2003, \mn@doi [\aj] {10.1086/342935}, \href
  {https://ui.adsabs.harvard.edu/abs/2003AJ....125.2348B} {125, 2348}

\bibitem[\protect\citeauthoryear{{Bonne}, {Brown}, {Jones}  \&
  {Pimbblet}}{{Bonne} et~al.}{2015}]{Bonne2015}
{Bonne} N.~J.,  {Brown} M. J.~I.,  {Jones} H.,   {Pimbblet} K.~A.,  2015,
  \mn@doi [\apj] {10.1088/0004-637X/799/2/160}, \href
  {https://ui.adsabs.harvard.edu/abs/2015ApJ...799..160B} {799, 160}

\bibitem[\protect\citeauthoryear{{Brammer} et~al.,}{{Brammer}
  et~al.}{2009}]{Brammer2009}
{Brammer} G.~B.,  et~al., 2009, \mn@doi [\apjl] {10.1088/0004-637X/706/1/L173},
  \href {https://ui.adsabs.harvard.edu/abs/2009ApJ...706L.173B} {706, L173}

\bibitem[\protect\citeauthoryear{{Bundy} et~al.,}{{Bundy}
  et~al.}{2010}]{Bundy2010}
{Bundy} K.,  et~al., 2010, \mn@doi [\apj] {10.1088/0004-637X/719/2/1969}, \href
  {https://ui.adsabs.harvard.edu/abs/2010ApJ...719.1969B} {719, 1969}

\bibitem[\protect\citeauthoryear{{Cardelli}, {Clayton}  \& {Mathis}}{{Cardelli}
  et~al.}{1989}]{Cardelli1989}
{Cardelli} J.~A.,  {Clayton} G.~C.,   {Mathis} J.~S.,  1989, \mn@doi [\apj]
  {10.1086/167900}, \href
  {https://ui.adsabs.harvard.edu/abs/1989ApJ...345..245C} {345, 245}

\bibitem[\protect\citeauthoryear{{Chilingarian} \& {Zolotukhin}}{{Chilingarian}
  \& {Zolotukhin}}{2012}]{Chilingarian2012}
{Chilingarian} I.~V.,  {Zolotukhin} I.~Y.,  2012, \mn@doi [\mnras]
  {10.1111/j.1365-2966.2011.19837.x}, \href
  {https://ui.adsabs.harvard.edu/abs/2012MNRAS.419.1727C} {419, 1727}

\bibitem[\protect\citeauthoryear{{Cooper} et~al.,}{{Cooper}
  et~al.}{2010}]{Cooper2010}
{Cooper} M.~C.,  et~al., 2010, \mn@doi [\mnras]
  {10.1111/j.1365-2966.2010.17312.x}, \href
  {https://ui.adsabs.harvard.edu/abs/2010MNRAS.409..337C} {409, 337}

\bibitem[\protect\citeauthoryear{{Cortese}}{{Cortese}}{2012}]{Cortese2012}
{Cortese} L.,  2012, \mn@doi [\aap] {10.1051/0004-6361/201219443}, \href
  {https://ui.adsabs.harvard.edu/abs/2012A&A...543A.132C} {543, A132}

\bibitem[\protect\citeauthoryear{{Dawson} et~al.,}{{Dawson}
  et~al.}{2013}]{Dawson2013}
{Dawson} K.~S.,  et~al., 2013, \mn@doi [\aj] {10.1088/0004-6256/145/1/10},
  \href {https://ui.adsabs.harvard.edu/abs/2013AJ....145...10D} {145, 10}

\bibitem[\protect\citeauthoryear{{Dieleman}, {Willett}  \& {Dambre}}{{Dieleman}
  et~al.}{2015}]{Dieleman2015}
{Dieleman} S.,  {Willett} K.~W.,   {Dambre} J.,  2015, \mn@doi [\mnras]
  {10.1093/mnras/stv632}, \href
  {https://ui.adsabs.harvard.edu/abs/2015MNRAS.450.1441D} {450, 1441}

\bibitem[\protect\citeauthoryear{{Dom{\'{\i}}nguez S{\'a}nchez},
  {Huertas-Company}, {Bernardi}, {Tuccillo}  \& {Fischer}}{{Dom{\'{\i}}nguez
  S{\'a}nchez} et~al.}{2018}]{dominguezsanchez2018}
{Dom{\'{\i}}nguez S{\'a}nchez} H.,  {Huertas-Company} M.,  {Bernardi} M.,
  {Tuccillo} D.,   {Fischer} J.~L.,  2018, \mn@doi [\mnras]
  {10.1093/mnras/sty338}, \href
  {http://adsabs.harvard.edu/abs/2018MNRAS.476.3661D} {476, 3661}

\bibitem[\protect\citeauthoryear{{Dom{\'\i}nguez S{\'a}nchez}
  et~al.,}{{Dom{\'\i}nguez S{\'a}nchez} et~al.}{2019}]{dominguezsanchez2019}
{Dom{\'\i}nguez S{\'a}nchez} H.,  et~al., 2019, \mn@doi [\mnras]
  {10.1093/mnras/sty3497}, \href
  {https://ui.adsabs.harvard.edu/abs/2019MNRAS.484...93D} {484, 93}

\bibitem[\protect\citeauthoryear{{Faber} et~al.,}{{Faber}
  et~al.}{2007}]{Faber2007}
{Faber} S.~M.,  et~al., 2007, \mn@doi [\apj] {10.1086/519294}, \href
  {https://ui.adsabs.harvard.edu/abs/2007ApJ...665..265F} {665, 265}

\bibitem[\protect\citeauthoryear{{Fang} et~al.,}{{Fang}
  et~al.}{2018}]{Fang2018}
{Fang} J.~J.,  et~al., 2018, \mn@doi [\apj] {10.3847/1538-4357/aabcba}, \href
  {https://ui.adsabs.harvard.edu/abs/2018ApJ...858..100F} {858, 100}

\bibitem[\protect\citeauthoryear{{Fraser-McKelvie}, {Brown}, {Pimbblet},
  {Dolley}, {Crossett}  \& {Bonne}}{{Fraser-McKelvie}
  et~al.}{2016}]{Fraser-McKelvie2016}
{Fraser-McKelvie} A.,  {Brown} M. J.~I.,  {Pimbblet} K.~A.,  {Dolley} T.,
  {Crossett} J.~P.,   {Bonne} N.~J.,  2016, \mn@doi [\mnras]
  {10.1093/mnrasl/slw117}, \href
  {https://ui.adsabs.harvard.edu/abs/2016MNRAS.462L..11F} {462, L11}

\bibitem[\protect\citeauthoryear{{Fraser-McKelvie}, {Brown}, {Pimbblet},
  {Dolley}  \& {Bonne}}{{Fraser-McKelvie} et~al.}{2018}]{Fraser-McKelvie2018}
{Fraser-McKelvie} A.,  {Brown} M. J.~I.,  {Pimbblet} K.,  {Dolley} T.,
  {Bonne} N.~J.,  2018, \mn@doi [\mnras] {10.1093/mnras/stx2823}, \href
  {https://ui.adsabs.harvard.edu/abs/2018MNRAS.474.1909F} {474, 1909}

\bibitem[\protect\citeauthoryear{{Hart} et~al.,}{{Hart}
  et~al.}{2016}]{Hart2016}
{Hart} R.~E.,  et~al., 2016, \mn@doi [\mnras] {10.1093/mnras/stw1588}, \href
  {https://ui.adsabs.harvard.edu/abs/2016MNRAS.461.3663H} {461, 3663}

\bibitem[\protect\citeauthoryear{{Hill} et~al.,}{{Hill}
  et~al.}{2011}]{Hill2011}
{Hill} D.~T.,  et~al., 2011, \mn@doi [\mnras]
  {10.1111/j.1365-2966.2010.17950.x}, \href
  {https://ui.adsabs.harvard.edu/abs/2011MNRAS.412..765H} {412, 765}

\bibitem[\protect\citeauthoryear{{Hocking}, {Geach}, {Davey}  \&
  {Sun}}{{Hocking} et~al.}{2015}]{hocking15}
{Hocking} A.,  {Geach} J.~E.,  {Davey} N.,   {Sun} Y.,  2015, arXiv e-prints,
  \href {https://ui.adsabs.harvard.edu/abs/2015arXiv150701589H} {p.
  arXiv:1507.01589}

\bibitem[\protect\citeauthoryear{{Hogg} et~al.,}{{Hogg}
  et~al.}{2002}]{Hogg2002}
{Hogg} D.~W.,  et~al., 2002, \mn@doi [\aj] {10.1086/341392}, \href
  {https://ui.adsabs.harvard.edu/abs/2002AJ....124..646H} {124, 646}

\bibitem[\protect\citeauthoryear{{Huertas-Company}, {Aguerri}, {Bernardi},
  {Mei}  \& {S{\'a}nchez Almeida}}{{Huertas-Company}
  et~al.}{2011}]{Huertas2011}
{Huertas-Company} M.,  {Aguerri} J.~A.~L.,  {Bernardi} M.,  {Mei} S.,
  {S{\'a}nchez Almeida} J.,  2011, \mn@doi [\aap]
  {10.1051/0004-6361/201015735}, \href
  {https://ui.adsabs.harvard.edu/abs/2011A&A...525A.157H} {525, A157}

\bibitem[\protect\citeauthoryear{{Huertas-Company} et~al.,}{{Huertas-Company}
  et~al.}{2015}]{Huertas2015}
{Huertas-Company} M.,  et~al., 2015, \mn@doi [\apjs]
  {10.1088/0067-0049/221/1/8}, \href
  {https://ui.adsabs.harvard.edu/abs/2015ApJS..221....8H} {221, 8}

\bibitem[\protect\citeauthoryear{{Jarrett}, {Chester}, {Cutri}, {Schneider},
  {Skrutskie}  \& {Huchra}}{{Jarrett} et~al.}{2000}]{Jarrett2000}
{Jarrett} T.~H.,  {Chester} T.,  {Cutri} R.,  {Schneider} S.,  {Skrutskie} M.,
   {Huchra} J.~P.,  2000, \mn@doi [\aj] {10.1086/301330}, \href
  {https://ui.adsabs.harvard.edu/abs/2000AJ....119.2498J} {119, 2498}

\bibitem[\protect\citeauthoryear{{Jarrett} et~al.,}{{Jarrett}
  et~al.}{2017}]{Jarrett2017}
{Jarrett} T.~H.,  et~al., 2017, \mn@doi [\apj] {10.3847/1538-4357/836/2/182},
  \href {https://ui.adsabs.harvard.edu/abs/2017ApJ...836..182J} {836, 182}

\bibitem[\protect\citeauthoryear{{Kruk} et~al.,}{{Kruk}
  et~al.}{2018}]{Kruk2018}
{Kruk} S.~J.,  et~al., 2018, \mn@doi [\mnras] {10.1093/mnras/stx2605}, \href
  {https://ui.adsabs.harvard.edu/abs/2018MNRAS.473.4731K} {473, 4731}

\bibitem[\protect\citeauthoryear{{Lange} et~al.,}{{Lange}
  et~al.}{2015}]{Lange2015}
{Lange} R.,  et~al., 2015, \mn@doi [\mnras] {10.1093/mnras/stu2467}, \href
  {https://ui.adsabs.harvard.edu/abs/2015MNRAS.447.2603L} {447, 2603}

\bibitem[\protect\citeauthoryear{{Lintott} et~al.,}{{Lintott}
  et~al.}{2008}]{Lintott2008}
{Lintott} C.~J.,  et~al., 2008, \mn@doi [\mnras]
  {10.1111/j.1365-2966.2008.13689.x}, \href
  {https://ui.adsabs.harvard.edu/abs/2008MNRAS.389.1179L} {389, 1179}

\bibitem[\protect\citeauthoryear{{Lotz}, {Primack}  \& {Madau}}{{Lotz}
  et~al.}{2004}]{Lotz2004}
{Lotz} J.~M.,  {Primack} J.,   {Madau} P.,  2004, \mn@doi [\aj]
  {10.1086/421849}, \href
  {https://ui.adsabs.harvard.edu/abs/2004AJ....128..163L} {128, 163}

\bibitem[\protect\citeauthoryear{{Mahajan} et~al.,}{{Mahajan}
  et~al.}{2020}]{Mahajan2020}
{Mahajan} S.,  et~al., 2020, \mn@doi [\mnras] {10.1093/mnras/stz2993}, \href
  {https://ui.adsabs.harvard.edu/abs/2020MNRAS.491..398M} {491, 398}

\bibitem[\protect\citeauthoryear{{Martin} et~al.,}{{Martin}
  et~al.}{2005}]{galex_paper}
{Martin} D.~C.,  et~al., 2005, \mn@doi [\apjl] {10.1086/426387}, \href
  {http://adsabs.harvard.edu/abs/2005ApJ...619L...1M} {619, L1}

\bibitem[\protect\citeauthoryear{{Masters} et~al.,}{{Masters}
  et~al.}{2010a}]{Masters2010dust}
{Masters} K.~L.,  et~al., 2010a, \mn@doi [\mnras]
  {10.1111/j.1365-2966.2010.16335.x}, \href
  {https://ui.adsabs.harvard.edu/abs/2010MNRAS.404..792M} {404, 792}

\bibitem[\protect\citeauthoryear{{Masters} et~al.,}{{Masters}
  et~al.}{2010b}]{Masters2010red}
{Masters} K.~L.,  et~al., 2010b, \mn@doi [\mnras]
  {10.1111/j.1365-2966.2010.16503.x}, \href
  {https://ui.adsabs.harvard.edu/abs/2010MNRAS.405..783M} {405, 783}

\bibitem[\protect\citeauthoryear{{Masters} et~al.,}{{Masters}
  et~al.}{2011}]{Masters2011bars}
{Masters} K.~L.,  et~al., 2011, \mn@doi [\mnras]
  {10.1111/j.1365-2966.2010.17834.x}, \href
  {https://ui.adsabs.harvard.edu/abs/2011MNRAS.411.2026M} {411, 2026}

\bibitem[\protect\citeauthoryear{{Masters} et~al.,}{{Masters}
  et~al.}{2019}]{Masters2019spirals}
{Masters} K.~L.,  et~al., 2019, \mn@doi [\mnras] {10.1093/mnras/stz1153}, \href
  {https://ui.adsabs.harvard.edu/abs/2019MNRAS.487.1808M} {487, 1808}

\bibitem[\protect\citeauthoryear{{Muzzin} et~al.,}{{Muzzin}
  et~al.}{2013}]{Muzzin2013}
{Muzzin} A.,  et~al., 2013, \mn@doi [\apjs] {10.1088/0067-0049/206/1/8}, \href
  {https://ui.adsabs.harvard.edu/abs/2013ApJS..206....8M} {206, 8}

\bibitem[\protect\citeauthoryear{{Nair} \& {Abraham}}{{Nair} \&
  {Abraham}}{2010}]{nair2010a}
{Nair} P.~B.,  {Abraham} R.~G.,  2010, \mn@doi [\apjs]
  {10.1088/0067-0049/186/2/427}, \href
  {http://adsabs.harvard.edu/abs/2010ApJS..186..427N} {186, 427}

\bibitem[\protect\citeauthoryear{{Nair}, {van den Bergh}  \& {Abraham}}{{Nair}
  et~al.}{2010}]{Nair2010b}
{Nair} P.~B.,  {van den Bergh} S.,   {Abraham} R.~G.,  2010, \mn@doi [\apj]
  {10.1088/0004-637X/715/1/606}, \href
  {https://ui.adsabs.harvard.edu/abs/2010ApJ...715..606N} {715, 606}

\bibitem[\protect\citeauthoryear{{Oh}, {Sarzi}, {Schawinski}  \& {Yi}}{{Oh}
  et~al.}{2011}]{Oh2011}
{Oh} K.,  {Sarzi} M.,  {Schawinski} K.,   {Yi} S.~K.,  2011, \mn@doi [\apjs]
  {10.1088/0067-0049/195/2/13}, \href
  {https://ui.adsabs.harvard.edu/abs/2011ApJS..195...13O} {195, 13}

\bibitem[\protect\citeauthoryear{{Otter}, {Masters}, {Simmons}  \&
  {Lintott}}{{Otter} et~al.}{2020}]{Otter2020}
{Otter} J.~A.,  {Masters} K.~L.,  {Simmons} B.,   {Lintott} C.~J.,  2020,
  \mn@doi [\mnras] {10.1093/mnras/stz3626}, \href
  {https://ui.adsabs.harvard.edu/abs/2020MNRAS.492.2722O} {492, 2722}

\bibitem[\protect\citeauthoryear{{Park} \& {Choi}}{{Park} \&
  {Choi}}{2005}]{ParkChoi2005}
{Park} C.,  {Choi} Y.-Y.,  2005, \mn@doi [\apjl] {10.1086/499243}, \href
  {https://ui.adsabs.harvard.edu/abs/2005ApJ...635L..29P} {635, L29}

\bibitem[\protect\citeauthoryear{{Patel}, {Holden}, {Kelson}, {Franx}, {van der
  Wel}  \& {Illingworth}}{{Patel} et~al.}{2012}]{Patel2012}
{Patel} S.~G.,  {Holden} B.~P.,  {Kelson} D.~D.,  {Franx} M.,  {van der Wel}
  A.,   {Illingworth} G.~D.,  2012, \mn@doi [\apjl]
  {10.1088/2041-8205/748/2/L27}, \href
  {https://ui.adsabs.harvard.edu/abs/2012ApJ...748L..27P} {748, L27}

\bibitem[\protect\citeauthoryear{{Planck Collaboration} et~al.,}{{Planck
  Collaboration} et~al.}{2016}]{planck16}
{Planck Collaboration} et~al., 2016, \mn@doi [\aap]
  {10.1051/0004-6361/201525830}, \href
  {https://ui.adsabs.harvard.edu/abs/2016A&A...594A..13P} {594, A13}

\bibitem[\protect\citeauthoryear{{Prescott} et~al.,}{{Prescott}
  et~al.}{2011}]{Prescott2011}
{Prescott} M.,  et~al., 2011, \mn@doi [\mnras]
  {10.1111/j.1365-2966.2011.19353.x}, \href
  {https://ui.adsabs.harvard.edu/abs/2011MNRAS.417.1374P} {417, 1374}

\bibitem[\protect\citeauthoryear{{Rowlands} et~al.,}{{Rowlands}
  et~al.}{2012}]{Rowlands2012}
{Rowlands} K.,  et~al., 2012, \mn@doi [\mnras]
  {10.1111/j.1365-2966.2011.19905.x}, \href
  {https://ui.adsabs.harvard.edu/abs/2012MNRAS.419.2545R} {419, 2545}

\bibitem[\protect\citeauthoryear{{Salim} et~al.,}{{Salim}
  et~al.}{2016}]{salim16}
{Salim} S.,  et~al., 2016, \mn@doi [\apjs] {10.3847/0067-0049/227/1/2}, \href
  {https://ui.adsabs.harvard.edu/abs/2016ApJS..227....2S} {227, 2}

\bibitem[\protect\citeauthoryear{{Schawinski} et~al.,}{{Schawinski}
  et~al.}{2009}]{Schawinski2009}
{Schawinski} K.,  et~al., 2009, \mn@doi [\mnras]
  {10.1111/j.1365-2966.2009.14793.x}, \href
  {https://ui.adsabs.harvard.edu/abs/2009MNRAS.396..818S} {396, 818}

\bibitem[\protect\citeauthoryear{{Schawinski} et~al.,}{{Schawinski}
  et~al.}{2014}]{Schawinski2014}
{Schawinski} K.,  et~al., 2014, \mn@doi [\mnras] {10.1093/mnras/stu327}, \href
  {https://ui.adsabs.harvard.edu/abs/2014MNRAS.440..889S} {440, 889}

\bibitem[\protect\citeauthoryear{{Schombert}}{{Schombert}}{2016}]{Schombert2016}
{Schombert} J.~M.,  2016, \mn@doi [\aj] {10.3847/0004-6256/152/6/214}, \href
  {https://ui.adsabs.harvard.edu/abs/2016AJ....152..214S} {152, 214}

\bibitem[\protect\citeauthoryear{{Schreiber} et~al.,}{{Schreiber}
  et~al.}{2018}]{Schreiber2018}
{Schreiber} C.,  et~al., 2018, \mn@doi [\aap] {10.1051/0004-6361/201833070},
  \href {https://ui.adsabs.harvard.edu/abs/2018A&A...618A..85S} {618, A85}

\bibitem[\protect\citeauthoryear{{Simard}, {Mendel}, {Patton}, {Ellison}  \&
  {McConnachie}}{{Simard} et~al.}{2011}]{Simard2011}
{Simard} L.,  {Mendel} J.~T.,  {Patton} D.~R.,  {Ellison} S.~L.,
  {McConnachie} A.~W.,  2011, \mn@doi [\apjs] {10.1088/0067-0049/196/1/11},
  \href {https://ui.adsabs.harvard.edu/abs/2011ApJS..196...11S} {196, 11}

\bibitem[\protect\citeauthoryear{{Simmons} \& {Urry}}{{Simmons} \&
  {Urry}}{2008}]{Simmons2008}
{Simmons} B.~D.,  {Urry} C.~M.,  2008, \mn@doi [\apj] {10.1086/589827}, \href
  {https://ui.adsabs.harvard.edu/abs/2008ApJ...683..644S} {683, 644}

\bibitem[\protect\citeauthoryear{{Simmons} et~al.,}{{Simmons}
  et~al.}{2017}]{Simmons2017}
{Simmons} B.~D.,  et~al., 2017, \mn@doi [\mnras] {10.1093/mnras/stw2587}, \href
  {https://ui.adsabs.harvard.edu/abs/2017MNRAS.464.4420S} {464, 4420}

\bibitem[\protect\citeauthoryear{{Skibba} et~al.,}{{Skibba}
  et~al.}{2009}]{Skibba2009}
{Skibba} R.~A.,  et~al., 2009, \mn@doi [\mnras]
  {10.1111/j.1365-2966.2009.15334.x}, \href
  {https://ui.adsabs.harvard.edu/abs/2009MNRAS.399..966S} {399, 966}

\bibitem[\protect\citeauthoryear{{Skrutskie} et~al.,}{{Skrutskie}
  et~al.}{2006}]{2mass_paper}
{Skrutskie} M.~F.,  et~al., 2006, \mn@doi [\aj] {10.1086/498708}, \href
  {https://ui.adsabs.harvard.edu/abs/2006AJ....131.1163S} {131, 1163}

\bibitem[\protect\citeauthoryear{{Smethurst} et~al.,}{{Smethurst}
  et~al.}{2015}]{Smethurst2015}
{Smethurst} R.~J.,  et~al., 2015, \mn@doi [\mnras] {10.1093/mnras/stv161},
  \href {https://ui.adsabs.harvard.edu/abs/2015MNRAS.450..435S} {450, 435}

\bibitem[\protect\citeauthoryear{{Steinmetz} \& {Navarro}}{{Steinmetz} \&
  {Navarro}}{2002}]{SteinmetzNavarro2002}
{Steinmetz} M.,  {Navarro} J.~F.,  2002, \mn@doi [\na]
  {10.1016/S1384-1076(02)00102-1}, \href
  {https://ui.adsabs.harvard.edu/abs/2002NewA....7..155S} {7, 155}

\bibitem[\protect\citeauthoryear{{Storrie-Lombardi}, {Lahav}, {Sodre}  \&
  {Storrie-Lombardi}}{{Storrie-Lombardi} et~al.}{1992}]{storrie92}
{Storrie-Lombardi} M.~C.,  {Lahav} O.,  {Sodre} L. J.,   {Storrie-Lombardi}
  L.~J.,  1992, \mn@doi [\mnras] {10.1093/mnras/259.1.8P}, \href
  {https://ui.adsabs.harvard.edu/abs/1992MNRAS.259P...8S} {259, 8P}

\bibitem[\protect\citeauthoryear{{Strateva} et~al.,}{{Strateva}
  et~al.}{2001}]{Strateva2001}
{Strateva} I.,  et~al., 2001, \mn@doi [\aj] {10.1086/323301}, \href
  {https://ui.adsabs.harvard.edu/abs/2001AJ....122.1861S} {122, 1861}

\bibitem[\protect\citeauthoryear{{Strauss} et~al.,}{{Strauss}
  et~al.}{2002}]{Strauss2002}
{Strauss} M.~A.,  et~al., 2002, \mn@doi [\aj] {10.1086/342343}, \href
  {http://adsabs.harvard.edu/abs/2002AJ....124.1810S} {124, 1810}

\bibitem[\protect\citeauthoryear{{Tohill}, {Ferreira}, {Conselice}, {Bamford}
  \& {Ferrari}}{{Tohill} et~al.}{2021}]{Tohill2021}
{Tohill} C.,  {Ferreira} L.,  {Conselice} C.~J.,  {Bamford} S.~P.,   {Ferrari}
  F.,  2021, \mn@doi [\apj] {10.3847/1538-4357/ac033c}, \href
  {https://ui.adsabs.harvard.edu/abs/2021ApJ...916....4T} {916, 4}

\bibitem[\protect\citeauthoryear{{Tojeiro} et~al.,}{{Tojeiro}
  et~al.}{2013}]{Tojeiro2013}
{Tojeiro} R.,  et~al., 2013, \mn@doi [\mnras] {10.1093/mnras/stt484}, \href
  {https://ui.adsabs.harvard.edu/abs/2013MNRAS.432..359T} {432, 359}

\bibitem[\protect\citeauthoryear{{Tuttle} \& {Tonnesen}}{{Tuttle} \&
  {Tonnesen}}{2020}]{Tuttle2020}
{Tuttle} S.~E.,  {Tonnesen} S.,  2020, \mn@doi [\apj]
  {10.3847/1538-4357/ab5dbb}, \href
  {https://ui.adsabs.harvard.edu/abs/2020ApJ...889..188T} {889, 188}

\bibitem[\protect\citeauthoryear{{Vavilova}, {Dobrycheva}, {Vasylenko},
  {Elyiv}, {Melnyk}  \& {Khramtsov}}{{Vavilova} et~al.}{2021}]{Vavilova2021}
{Vavilova} I.~B.,  {Dobrycheva} D.~V.,  {Vasylenko} M.~Y.,  {Elyiv} A.~A.,
  {Melnyk} O.~V.,   {Khramtsov} V.,  2021, \mn@doi [\aap]
  {10.1051/0004-6361/202038981}, \href
  {https://ui.adsabs.harvard.edu/abs/2021A&A...648A.122V} {648, A122}

\bibitem[\protect\citeauthoryear{{Vika}, {Vulcani}, {Bamford},
  {H{\"a}u{\ss}ler}  \& {Rojas}}{{Vika} et~al.}{2015}]{Vika2015}
{Vika} M.,  {Vulcani} B.,  {Bamford} S.~P.,  {H{\"a}u{\ss}ler} B.,   {Rojas}
  A.~L.,  2015, \mn@doi [\aap] {10.1051/0004-6361/201425174}, \href
  {https://ui.adsabs.harvard.edu/abs/2015A&A...577A..97V} {577, A97}

\bibitem[\protect\citeauthoryear{{Walmsley} et~al.,}{{Walmsley}
  et~al.}{2020}]{Walmsley2020}
{Walmsley} M.,  et~al., 2020, \mn@doi [\mnras] {10.1093/mnras/stz2816}, \href
  {https://ui.adsabs.harvard.edu/abs/2020MNRAS.491.1554W} {491, 1554}

\bibitem[\protect\citeauthoryear{{Walmsley} et~al.,}{{Walmsley}
  et~al.}{2021}]{Walmsley2021}
{Walmsley} M.,  et~al., 2021, arXiv e-prints, \href
  {https://ui.adsabs.harvard.edu/abs/2021arXiv210208414W} {p. arXiv:2102.08414}

\bibitem[\protect\citeauthoryear{{Weinmann}, {van den Bosch}, {Yang}  \&
  {Mo}}{{Weinmann} et~al.}{2006}]{Weinmann2006}
{Weinmann} S.~M.,  {van den Bosch} F.~C.,  {Yang} X.,   {Mo} H.~J.,  2006,
  \mn@doi [\mnras] {10.1111/j.1365-2966.2005.09865.x}, \href
  {https://ui.adsabs.harvard.edu/abs/2006MNRAS.366....2W} {366, 2}

\bibitem[\protect\citeauthoryear{{Willett} et~al.,}{{Willett}
  et~al.}{2013}]{Willett2013}
{Willett} K.~W.,  et~al., 2013, \mn@doi [\mnras] {10.1093/mnras/stt1458}, \href
  {https://ui.adsabs.harvard.edu/abs/2013MNRAS.435.2835W} {435, 2835}

\bibitem[\protect\citeauthoryear{{Willett} et~al.,}{{Willett}
  et~al.}{2017}]{Willett2017}
{Willett} K.~W.,  et~al., 2017, \mn@doi [\mnras] {10.1093/mnras/stw2568}, \href
  {https://ui.adsabs.harvard.edu/abs/2017MNRAS.464.4176W} {464, 4176}

\bibitem[\protect\citeauthoryear{{Willmer} et~al.,}{{Willmer}
  et~al.}{2006}]{Willmer2006}
{Willmer} C.~N.~A.,  et~al., 2006, \mn@doi [\apj] {10.1086/505455}, \href
  {https://ui.adsabs.harvard.edu/abs/2006ApJ...647..853W} {647, 853}

\bibitem[\protect\citeauthoryear{{Wolf} et~al.,}{{Wolf}
  et~al.}{2009}]{Wolf2009}
{Wolf} C.,  et~al., 2009, \mn@doi [\mnras] {10.1111/j.1365-2966.2008.14204.x},
  \href {https://ui.adsabs.harvard.edu/abs/2009MNRAS.393.1302W} {393, 1302}

\bibitem[\protect\citeauthoryear{{Wright} et~al.,}{{Wright}
  et~al.}{2010}]{Wright2010}
{Wright} E.~L.,  et~al., 2010, \mn@doi [\aj] {10.1088/0004-6256/140/6/1868},
  \href {https://ui.adsabs.harvard.edu/abs/2010AJ....140.1868W} {140, 1868}

\bibitem[\protect\citeauthoryear{{Xu, K.}, {Liu}, {Jing}, {Sawicki}  \&
  {Gwyn}}{{Xu, K.} et~al.}{2021}]{Xu2021}
{Xu, K.} K.,  {Liu} C.,  {Jing} Y.,  {Sawicki} M.,   {Gwyn} S.,  2021, \mn@doi
  [Science China Physics, Mechanics, and Astronomy]
  {10.1007/s11433-020-1667-0}, \href
  {https://ui.adsabs.harvard.edu/abs/2021SCPMA..6479811X} {64, 279811}

\bibitem[\protect\citeauthoryear{{Zehavi} et~al.,}{{Zehavi}
  et~al.}{2011}]{Zehavi2011}
{Zehavi} I.,  et~al., 2011, \mn@doi [\apj] {10.1088/0004-637X/736/1/59}, \href
  {https://ui.adsabs.harvard.edu/abs/2011ApJ...736...59Z} {736, 59}

\bibitem[\protect\citeauthoryear{{Zwicky}}{{Zwicky}}{1955}]{Zwicky1955}
{Zwicky} F.,  1955, \mn@doi [\pasp] {10.1086/126807}, \href
  {https://ui.adsabs.harvard.edu/abs/1955PASP...67..232Z} {67, 232}

\bibitem[\protect\citeauthoryear{{van den Bergh}}{{van den
  Bergh}}{1976}]{vandenBergh1976}
{van den Bergh} S.,  1976, \mn@doi [\apj] {10.1086/154452}, \href
  {https://ui.adsabs.harvard.edu/abs/1976ApJ...206..883V} {206, 883}

\bibitem[\protect\citeauthoryear{{van den Bosch}, {Aquino}, {Yang}, {Mo},
  {Pasquali}, {McIntosh}, {Weinmann}  \& {Kang}}{{van den Bosch}
  et~al.}{2008}]{VandenBosch2008}
{van den Bosch} F.~C.,  {Aquino} D.,  {Yang} X.,  {Mo} H.~J.,  {Pasquali} A.,
  {McIntosh} D.~H.,  {Weinmann} S.~M.,   {Kang} X.,  2008, \mn@doi [\mnras]
  {10.1111/j.1365-2966.2008.13230.x}, \href
  {https://ui.adsabs.harvard.edu/abs/2008MNRAS.387...79V} {387, 79}

\makeatother
\end{thebibliography}








\bsp	
\label{lastpage}
\end{document}